\documentclass[12pt]{article}

\usepackage{sbc-template}

\usepackage[english]{babel}   
\usepackage[T1]{fontenc} 
\usepackage[utf8]{inputenc} 
\usepackage{multicol}
\usepackage{graphicx,url}
\usepackage{amsmath,amssymb,amsfonts}
\usepackage{algorithmic}
\usepackage{textcomp}
\usepackage{enumerate}
\usepackage{multirow}
\usepackage{soul}
\usepackage{verbatim}
\usepackage[table,xcdraw]{xcolor}
\usepackage[shortlabels]{enumitem}
\usepackage{scalefnt} 
\usepackage{comment} 
\usepackage{lscape}
\usepackage{fancyhdr}
\usepackage{pdfcomment} 
\usepackage{acronym}
\usepackage{tabularx}
\usepackage{array}
\usepackage{placeins}
\usepackage{longtable}
\usepackage{orcidlink}
\hypersetup{
  colorlinks=false,
  pdfborder={0 0 0},
  pdfhighlight=/N,
  linkbordercolor={1 1 1},
  citebordercolor={1 1 1},
  urlbordercolor={1 1 1},
  filebordercolor={1 1 1},
  menubordercolor={1 1 1},
  runbordercolor={1 1 1}
}
\makeatletter
\newcommand{\defautorano}[3]{%
  \expandafter\def\csname autor@#1\endcsname{#2}%
  \expandafter\def\csname ano@#1\endcsname{#3}%
}
\newcommand{\autorano}[1]{%
  \ifcsname autor@#1\endcsname
    \csname autor@#1\endcsname~(\csname ano@#1\endcsname)\nocite{#1}%
  \else
    \textbf{[author/year not mapped: #1]}\nocite{#1}%
  \fi
}
\makeatother

\newacro{HCI}{Human-Computer Interaction}
\newacro{TAM}{Technology Acceptance Model}
\newacro{TAM2}{Technology Acceptance Model 2}
\newacro{TAM3}{Technology Acceptance Model 3}
\newacro{UTAUT}{Unified Theory of Acceptance and Use of Technology}
\newacro{UTAUT2}{Unified Theory of Acceptance and Use of Technology 2}
\newacro{AI}{Artificial Intelligence}
\newacro{LLM}{Large Language Models}
\newacro{RQ}{Research Question}
\newacro{mHealth}{Mobile Health}
\newacro{TRA}{Theory of Reasoned Action}
\newacro{TPB}{Theory of Planned Behavior}
\newacro{IDT}{Diffusion of Innovations Theory}
\newacro{MM}{Motivational Model}
\newacro{TAM-TPB}{TAM-TPB}
\newacro{MPCU}{Model of PC Utilization}
\newacro{SCT}{Social Cognitive Theory}
\newacro{DSR}{Design Science Research}
\newacro{IS}{Information Systems}
\newacro{TRI}{Technology Readiness Index}
\newacro{SIGProj}{Project Information and Management System}
\defautorano{davis1989perceived}{Davis}{1989}
\defautorano{venkateshTheoreticalExtensionTechnology2000b}{Venkatesh and Davis}{2000}
\defautorano{venkateshTechnologyAcceptanceModel2008}{Venkatesh and Bala}{2008}
\defautorano{venkateshUserAcceptanceInformation2003b}{Venkatesh et al.}{2003}
\defautorano{venkateshConsumerAcceptanceUse2012a}{Venkatesh et al.}{2012}
\defautorano{webster2002analyzing}{Webster and Watson}{2002}
\defautorano{pinheiroAvaliacaoUsabilidadeSistema2023a}{Pinheiro et al.}{2023}
\defautorano{wu2024investigating}{Wu and Lim}{2024}
\defautorano{hong2014framework}{Hong et al.}{2014}
\defautorano{mkhomazi2013guide}{Mkhomazi and Iyamu}{2013}
\defautorano{brown2015technology}{Brown et al.}{2015}
\defautorano{taherdoost2018review}{Taherdoost}{2018}
\newcolumntype{Y}{>{\raggedright\arraybackslash}X}

\newcommand{\antesquadro}{\vspace{0.6\baselineskip}}

\newcommand{\fontequadro}[1]{%
  \vspace{0.25\baselineskip}%
  \begin{center}%
    \parbox{0.95\linewidth}{\centering\scriptsize\textit{Source:} #1}%
  \end{center}%
  \vspace{0.6\baselineskip}%
}

\newif\ifblindreview
\blindreviewfalse %colocar false ativar as informações.

\newlength{\blindreviewtitletopspace}
\newlength{\blindreviewtitleabstractspace}
\newcommand{\paperauthorfont}{\normalfont\large\itshape}
\newcommand{\paperaddressfont}{\normalfont\normalsize}
\newcommand{\paperemailfont}{\normalfont\footnotesize\ttfamily}
\newcommand{\orcidlinkauthor}[1]{\raisebox{-0.05ex}{\scalebox{1.25}{\orcidlink{#1}}}}
\newcommand{\paperauthor}{Nathalino Pachêco Britto\,\orcidlinkauthor{0009-0004-9017-1643}\inst{1}}
\newcommand{\paperaddress}{Doctoral Program in Computer Science (DCCMAPI) \\ -- UFMA/UFPI Association -- Federal University of Piauí
  (UFPI)\\
  ZIP Code: 64049-550 -- Teresina  -- PI -- Brazil\\
  {\paperemailfont nathalino.britto@ufpi.edu.br}
}

\title{A Decision Framework for Selecting Technology Acceptance and Use Models: TAM, TAM2, TAM3, UTAUT, and UTAUT2}

\author{\paperauthor}

\address{\paperaddress}

\makeatletter
\ifblindreview
\def\@maketitle{\newpage
 \begin{center}
 \vspace*{\blindreviewtitletopspace}
 {\XVIPT\bf\@title\par}
 \vskip\blindreviewtitleabstractspace
 \end{center}\par
}
\else
\def\@maketitle{\newpage
 \begin{center}
 \vspace*{-.7cm}
 {\XVIPT\bf\@title\par}
 \vglue 7pt plus 3pt minus 3pt
 {\paperauthorfont
  \begin{tabular}[t]{c}\@author\end{tabular}\par}
 \vglue 7pt plus 3pt minus 3pt
 {\paperaddressfont
  \begin{tabular}[t]{c}\inst{\instnum}\,\@address\end{tabular}\par}
 \vglue 6pt plus 3pt minus 3pt
 \end{center}\par
}
\fi
\makeatother

\begin{document} 

\maketitle

\thispagestyle{plain}

\begin{abstract}
\textbf{Introduction:} Models such as TAM, TAM2, TAM3, UTAUT, and UTAUT2 underpin a substantial share of research on technology acceptance and use in HCI and related fields. Although they differ in terms of constructs and conditions of application, their selection is often not conceptually justified, frequently driven by pragmatic considerations, with implications for theoretical consistency and cross-study comparability. \textbf{Objective:} To propose a decision framework that guides the selection among the five models of the TAM/UTAUT lineage based on conceptual criteria derived from their structural differences. \textbf{Methods:} A conceptual, artifact-oriented approach was adopted, comprising comparative analysis of the models and critical review studies, derivation of conceptual dimensions, and formulation of operational decision criteria. Applicability was demonstrated through two contrasting research scenarios. \textbf{Results:} The framework articulates five analytical dimensions, 16 guiding questions, and descriptive suitability profiles for the models. The demonstration illustrated, in the two scenarios examined, its ability to identify misalignments between the declared model and the actual operationalization, as well as to confirm theoretical choices consistent with the conditions of the study.
\end{abstract}


\section{Introduction}\label{sec:introducao}

Understanding the factors that influence technology acceptance and use has become a central concern in \ac{HCI} and related fields, especially given the increasing incorporation of digital systems into social, organizational, and educational contexts. In this scenario, theoretical models such as \ac{TAM}, proposed by \autorano{davis1989perceived} in the late 1980s, and its subsequent extensions (\ac{TAM2} \cite{venkateshTheoreticalExtensionTechnology2000b} and \ac{TAM3} \cite{venkateshTechnologyAcceptanceModel2008}), as well as \ac{UTAUT}, proposed by \autorano{venkateshUserAcceptanceInformation2003b} in the early 2000s, and its evolution (\ac{UTAUT2} \cite{venkateshConsumerAcceptanceUse2012a}), constitute a theoretical lineage spanning more than three decades of development and refinement \cite{hornbaekTechnologyAcceptanceUser2017a, debritoLimitacoesDosModelos2019a, batista2023consideraccoes}.

The recurrence of these models in the literature is not episodic. Recent reviews confirm that such approaches continue to be widely applied across multiple domains, ranging from complex organizational systems \cite{chatterjee2023assessing} to emerging technologies such as \ac{mHealth} applications \cite{nadalTechnologyAcceptanceMobile2020a, leeUnderstandingPsychosocialBarriers2025a} and intelligent environments \cite{rico2020smart, fakhrhosseiniUserAdoptionIntelligent2024a}. Despite this widespread diffusion, the selection of these models in many studies is often not accompanied by an explicit statement of the conceptual criteria that justify it. Frequently, the adoption of \ac{TAM}, its extensions, or \ac{UTAUT} is presented implicitly, as can be inferred from the recurrent pattern of application of these approaches \cite{fakhrhosseiniUserAdoptionIntelligent2024a}.

Instead of being grounded in an analysis of structural differences, explanatory scope, and contextual suitability, the selection of these models tends to be driven by pragmatic considerations. As a consequence, relevant conceptual nuances tend to be disregarded, including the nature of the constructs involved, the explanatory mechanisms proposed, and the conditions under which each model was originally conceived and validated. This adoption pattern, widely observed in the literature, indicates not only the diffusion of these models but also limitations in the way they are selected and applied across different research contexts \cite{debritoLimitacoesDosModelos2019a, batista2023consideraccoes}.

This gap is not fully addressed by existing comparative studies. Although works such as those by \autorano{brown2015technology} and \autorano{taherdoost2018review} compare technology acceptance and use models, their analyses focus on aspects such as explanatory power or empirical performance, without explicitly structuring the decision process involved in choosing the most suitable model for a given context. In an exploratory search of the main Brazilian \ac{HCI} publication venues, such as the proceedings of the Brazilian Symposium on Human Factors in Computing Systems, the Journal on Interactive Systems, and the Journal of the Brazilian Computer Society, no critical reviews dedicated to these models that guide researchers in theoretical selection were identified. This absence is particularly relevant considering that the Brazilian \ac{HCI} community itself recognized, in the Grand Research Challenges for 2025--2035, the need to revisit and develop new theoretical and methodological foundations for the field \cite{pereira2024grandihc}.

The absence of approaches that organize conceptual criteria in a structured and operational manner to guide theoretical selection constitutes a methodological gap with direct impact on the theoretical consistency of studies, cross-study comparability, and the quality of the inferences produced. This scenario is intensified by the rapid pace of technological evolution, which continuously expands the application contexts of these models. The growing investigation of human factors in systems based on \ac{AI} \cite{silvaHumanFactorsDesign2024a} and the identification of acceptance models as theoretically influential approaches in contemporary \ac{HCI} research \cite{mohammadzadeh2025identifying} reinforce the need for explicit and replicable criteria to guide the selection of the most suitable model for the research context under investigation.

In light of the above, the following \ac{RQ} is defined: how can technology acceptance and use models be selected in a structured and conceptually grounded manner, considering different research contexts and objectives? To answer this question, this study proposes a decision framework that assists researchers in choosing the most suitable model for their research context, based on the identification of conceptual criteria derived from the literature.

The proposal is conceptual and methodological in nature, oriented toward the construction of a decision-support artifact \cite{hevner2004design}. Rather than introducing a new acceptance model or extending existing ones, the framework makes explicit the criteria that guide the selection among the five models of the \ac{TAM}/\ac{UTAUT} lineage, organizing them into five analytical dimensions: application context, intended type of use, adoption stage under consideration, predominant explanatory focus, and intended analytical granularity. These dimensions are operationalized through 16 guiding questions and complemented by descriptive suitability profiles for the models. This approach shifts the focus from performance-based comparison to the researcher's decision process, promoting greater transparency and traceability in the justification of theoretical choices.

The application of the framework is illustrated through two contrasting research scenarios, in which the artifact operates in two ways: through constructive divergence, by identifying that a study's operationalization was more compatible with a model different from the one declared by the authors; and through conceptually grounded confirmation, by supporting another study's theoretical choice and recognizing the coherence of adaptations made in light of the research context.

The paper is organized as follows. Section~\ref{sec:Fundamentos} presents the theoretical foundations of the analyzed models; Section~\ref{sec:AbordagemEstudo} describes the methodological approach; Section~\ref{sec:AnaliseComparativa} develops the comparative analysis; Section~\ref{sec:Framework} presents the proposed framework; and Section~\ref{sec:AplicacaoFrameworkDecisaoSelecao} demonstrates its application. Finally, Sections~\ref{sec:DiscussaoLimitacoes} and~\ref{sec:Conclusao} bring together the discussion, limitations, conclusions, and directions for future work.

\section{Theoretical Foundations of Technology Acceptance and Use Models} \label{sec:Fundamentos}

This section presents the theoretical foundations of the technology acceptance and use models analyzed in this study, situating them within a conceptual lineage that originates from antecedents in social psychology and the diffusion of innovations and evolves through successive expansions of explanatory scope. In this trajectory, the following models are considered: \ac{TAM} \cite{davis1989perceived}, \ac{TAM2} \cite{venkateshTheoreticalExtensionTechnology2000b}, \ac{UTAUT} \cite{venkateshUserAcceptanceInformation2003b}, \ac{TAM3} \cite{venkateshTechnologyAcceptanceModel2008}, and \ac{UTAUT2} \cite{venkateshConsumerAcceptanceUse2012a}. The presentation, however, follows a conceptual and evolutionary logic, prioritizing relationships of theoretical continuity and showing how each proposal preserves, modifies, or extends elements from previous formulations.

\subsection{Theoretical Antecedents of Acceptance Models} \label{sec:AntecedentesModelos}

Technology acceptance and use models have their roots in established theories from social psychology and the diffusion of innovations. The \ac{TRA} posits that human behavior is determined by behavioral intention, which results from individual attitudes and subjective norms \cite{fishbein1975foundations}. The \ac{TPB} extends this framework by introducing perceived behavioral control, recognizing that external factors can influence the performance of a behavior even when there is a favorable intention \cite{ajzen1991theory}. Complementarily, \ac{IDT} shifts the focus to the social context of adoption, emphasizing how characteristics of the innovation, such as relative advantage, compatibility, complexity, trialability, and observability, influence the speed and extent of adoption across different social systems \cite{rogersDiffusionInnovations1983}.

These antecedents constitute the conceptual foundation of the specific technology acceptance models. The approaches derived from \ac{TRA} and \ac{TPB} contributed to the understanding of the cognitive and intentional factors associated with use behavior, while \ac{IDT} introduced elements related to the characteristics of the technology and the collective context of adoption. The articulation of these perspectives fostered the development of progressively more comprehensive models, as is particularly evident in \ac{UTAUT}, which synthesizes multiple approaches into a unified framework \cite{venkateshUserAcceptanceInformation2003b}.

\subsection{Technology Acceptance Model (TAM)} \label{sec:TAM}

\ac{TAM}, proposed by \autorano{davis1989perceived}, is one of the most influential models in research on technology acceptance and use. Developed from foundations in social psychology, the model proposes a conceptually simple and parsimonious structure to explain how individuals come to accept and use technological systems based on their cognitive perceptions.

The model is structured around two core constructs: perceived usefulness, defined as the degree to which an individual believes that using a technology can improve their performance on a task; and perceived ease of use, defined as the degree to which the use of the technology is perceived as free of effort \cite{davis1989perceived}. In causal terms, perceived ease of use directly influences perceived usefulness, and both constructs influence behavioral intention, which in turn determines actual system use. It should be noted that the attitude construct, present in the original formulation, was subsequently removed due to its lack of significant predictive power, bringing \ac{TAM} closer to a more direct structure linking beliefs to intention \cite{venkateshTheoreticalExtensionTechnology2000b}.

\ac{TAM} is particularly suited to contexts in which user behavior can be explained predominantly by perceptions of usefulness and effort. Across different technological contexts, TAM-based specifications have reported explanatory power for behavioral intention ranging between 38\% and 53\% of the observed variance \cite{venkateshUserAcceptanceInformation2003b}.

\subsection{Technology Acceptance Model 2 (TAM2)} \label{sec:TAM2}

\ac{TAM2}, proposed by \autorano{venkateshTheoreticalExtensionTechnology2000b}, was developed to extend the explanatory power of \ac{TAM} by incorporating additional determinants of perceived usefulness, organized into two groups: social influence processes and cognitive instrumental processes related to the job.

Among the social influence processes, the model includes subjective norm, image, and voluntariness of use. Subjective norm acts as a direct determinant of behavioral intention in mandatory use contexts, operating through compliance, and as an indirect determinant of perceived usefulness through internalization and identification. Image, in turn, influences perceived usefulness by capturing the degree to which the use of the technology is associated with status enhancement. Among the cognitive instrumental processes, the model incorporates job relevance, output quality, and result demonstrability, all acting as determinants of perceived usefulness \cite{venkateshTheoreticalExtensionTechnology2000b}. Experience and voluntariness of use act as moderators of the relationships among constructs, altering the strength of the effects as a function of the user's level of familiarity and the mandatory or voluntary nature of use.

\ac{TAM2} extends the original model by incorporating social and job-related factors while maintaining the core structure based on perceived usefulness and perceived ease of use. However, the determinants of perceived ease of use had not yet been formally integrated into the TAM2 structure, motivating the extension proposed in \ac{TAM3}.

\subsection{Technology Acceptance Model 3 (TAM3)} \label{sec:TAM3}

\ac{TAM3}, proposed by \autorano{venkateshTechnologyAcceptanceModel2008}, represents an extension of \ac{TAM2} aimed at deepening the determinants of perceived ease of use, which in previous versions had received less detailed treatment of its specific antecedents.

The model organizes these determinants according to an anchor and adjustment framework. The anchors comprise computer self-efficacy, perceptions of external control, computer anxiety, and computer playfulness. The adjustments comprise perceived enjoyment and objective usability. In causal terms, these antecedents influence perceived ease of use, while the determinants of perceived usefulness remain linked to the structure inherited from \ac{TAM2}. The model keeps the antecedents of perceived usefulness and perceived ease of use separate, proposing no crossover effects between these two sets of determinants \cite{venkateshTechnologyAcceptanceModel2008}.

User experience acts as a moderator of the relationships between antecedents and perceived ease of use over time. As familiarity increases, the effects of some individual antecedents, such as anxiety and self-efficacy, tend to diminish, while factors associated with direct interaction with the system come to exert greater influence on use perceptions \cite{venkateshTechnologyAcceptanceModel2008}.

\ac{TAM3} represents a significant increase in structural complexity by integrating individual and contextual factors into a more granular structure \cite{debritoLimitacoesDosModelos2019a}. The characteristics that conceptually distinguish \ac{TAM3} from its predecessors will be revisited in the comparative analysis presented in Section~\ref{sec:AnaliseComparativa}. In parallel with these incremental extensions within the \ac{TAM} lineage, the field moved toward a distinct approach oriented to the integration of multiple theoretical traditions.

\subsection{Unified Theory of Acceptance and Use of Technology (UTAUT)} \label{sec:UTAUT}

\ac{UTAUT}, proposed by \autorano{venkateshUserAcceptanceInformation2003b}, represents a synthesis of eight technology acceptance models, including \ac{TAM}, \ac{TRA}, \ac{TPB}, and \ac{IDT}, as well as \ac{MM}, \ac{MPCU}, the combined \ac{TAM-TPB} model, and \ac{SCT}. Its development stemmed from the observation that researchers tended to select constructs from different models or to adopt a favored model while disregarding the contributions of others. \ac{UTAUT} proposes a unified framework to explain behavioral intention and use behavior of information technology \cite{venkateshUserAcceptanceInformation2003b}.

The model organizes four core constructs with distinct causal relationships. Performance expectancy, effort expectancy, and social influence act as direct determinants of behavioral intention. Facilitating conditions directly influence use behavior, without mediation through intention, when effort expectancy is already present in the model. Behavioral intention, in turn, determines use behavior \cite{venkateshUserAcceptanceInformation2003b, debritoLimitacoesDosModelos2019a}. Four moderating variables, gender, age, experience, and voluntariness of use, operate on the relationships among constructs, with distinct patterns for each association.

From an empirical standpoint, \ac{UTAUT} demonstrated explanatory power superior to that of previous models, accounting for approximately 70\% of the variance in behavioral intention \cite{venkateshUserAcceptanceInformation2003b}. This formulation, however, remained strongly anchored in the organizational context, motivating its extension to individual consumer settings.

\subsection{Unified Theory of Acceptance and Use of Technology 2 (UTAUT2)} \label{sec:UTAUT2}

\ac{UTAUT2}, proposed by \autorano{venkateshConsumerAcceptanceUse2012a}, represents an extension of \ac{UTAUT} oriented toward the individual consumer context, where technology use tends to be voluntary and influenced by factors distinct from those identified in organizational settings.

To address this specificity, \ac{UTAUT2} incorporates three additional constructs beyond the four already present in \ac{UTAUT}. Hedonic motivation, defined as the fun or pleasure derived from using the technology, plays an important role in the formation of behavioral intention in consumer contexts. Price value, defined as the cognitive trade-off between the perceived benefits and the monetary cost of use, also directly influences intention. Habit, defined as the degree to which an individual tends to perform a behavior automatically as a result of prior experience, influences both behavioral intention and use behavior directly \cite{venkateshConsumerAcceptanceUse2012a}. The moderating variables are gender, age, and experience, with the exclusion of voluntariness of use given the predominantly voluntary nature of the consumer context.

From an empirical standpoint, the extensions proposed in \ac{UTAUT2} resulted in a substantial increase in explanatory power: when \ac{UTAUT} was reestimated in the consumer context of the same study, the variance explained in behavioral intention rose from 56\% to 74\%, and the variance in use behavior increased from 40\% to 52\% \cite{venkateshConsumerAcceptanceUse2012a}.

\subsection{Conceptual Synthesis} \label{sec:SinteseConceitual}

The analysis of the technology acceptance and use models presented reveals a progressive evolution in how adoption behavior is understood. From initial approaches centered on individual cognitive determinants, as in \ac{TAM}, a gradual incorporation of social, contextual, and experiential factors can be observed in subsequent extensions, resulting in progressively more comprehensive approaches such as \ac{UTAUT} and \ac{UTAUT2}. This trajectory reflects a continuous movement in the literature toward broadening the explanatory scope of these approaches, seeking to capture the complexity inherent in human behavior regarding technology use \cite{debritoLimitacoesDosModelos2019a, batista2023consideraccoes}.

From this evolution, it is possible to identify common conceptual dimensions that structure the analyzed models, among which the nature of the determinants considered, the predominant level of analysis, and the degree of structural complexity are particularly relevant. These aspects make it clear that the models differ not only in terms of specific constructs but also in how they organize and interpret the phenomenon of technology acceptance \cite{debritoLimitacoesDosModelos2019a}. The structural differences associated with these aspects, as well as their implications for model selection in research, are examined in detail in Section~\ref{sec:AnaliseComparativa}.

\section{Study Approach}\label{sec:AbordagemEstudo}

This study adopts a conceptual approach oriented toward the construction of a decision-support artifact. The framework is built upon the analysis and synthesis of existing literature, without primary data collection during the criteria derivation stage. The illustrative application of the framework to research scenarios constitutes a complementary demonstration stage, consistent with \ac{DSR} principles \cite{hevner2004design}.

This choice is justified by the objective of the investigation: not to explain or predict use behavior, but to organize, in an explicit and traceable manner, criteria that can guide the selection of models already established in the literature. Approaches of this nature have precedent in \ac{IS} research, where conceptual synthesis aimed at framework construction is recognized as a legitimate form of theoretical contribution, provided that the derivation process is transparent and open to scrutiny \cite{whetten1989constitutes, hevner2004design, jabareen2009building}. The quality of the proposal, therefore, is assessed not exclusively by empirical criteria, but also by the internal coherence of the framework, the traceability of its criteria to the examined literature, and the potential usefulness of the proposal for guiding research decisions \cite{hevner2004design}.

The review corpus was delimited based on two inclusion criteria. The first restricted the selection to the foundational articles of the five analyzed models: \ac{TAM} \cite{davis1989perceived}, \ac{TAM2} \cite{venkateshTheoreticalExtensionTechnology2000b}, \ac{TAM3} \cite{venkateshTechnologyAcceptanceModel2008}, \ac{UTAUT} \cite{venkateshUserAcceptanceInformation2003b}, and \ac{UTAUT2} \cite{venkateshConsumerAcceptanceUse2012a}. These articles were included regardless of their empirical components, as it is in these works that the original authors make explicit the assumptions, constructs, moderators, target contexts, and boundary conditions that delimit the scope of each proposal.

The second criterion included critical review and comparative analysis studies, purposefully selected based on three functional subcriteria: (a) the study analyzes limitations or boundary conditions of the models rather than applying them to specific empirical contexts; (b) the scope of the work covers more than one model of the \ac{TAM} or \ac{UTAUT} lineage within the same publication; and (c) the study explicitly identifies conditions under which a given model presents limitations, requires adaptation, or has lower analytical suitability \cite{turner2010does, williams2015unified, debritoLimitacoesDosModelos2019a, malatji2020understanding, batista2023consideraccoes}.

Studies that apply the models to specific empirical contexts without contributing to their conceptual definition were excluded. The theoretical antecedents \ac{TRA} \cite{fishbein1975foundations}, \ac{TPB} \cite{ajzen1991theory}, and \ac{IDT} \cite{rogersDiffusionInnovations1983} were included to contextualize the conceptual lineage of the models, not as objects of comparative analysis. The resulting corpus comprises ten primary and critical review sources, complemented by three works of theoretical contextualization related to the antecedents of the models. Additional studies were drawn upon selectively based on their direct relevance to the analytical claims formulated, without being part of the extraction corpus \cite{dwivedi2019re, royApplyingUTAUT2Qualitative2025a, mohammadzadeh2025identifying, hong2014framework, mkhomazi2013guide}.

The analysis of the selected sources was conducted through systematic reading oriented toward the extraction of comparable theoretical elements. For each model, the following elements were recorded in individual extraction sheets: core constructs and their definitions, proposed causal relationships, moderating variables, original application context, assumed type of use, and primary adoption stage. The extraction sheets are available as supplementary material in a public repository \cite{anonymous_2026_19714800}. This procedure followed the logic of concept-driven reviews, in which the unit of analysis is the concept or dimension rather than the chronology of publications \cite{webster2002analyzing, okoli2015guide}.

The use of extraction sheets made it possible to identify, by comparing the records for each model, which aspects were recurrent in the lineage and which were specific to particular proposals. This material was subjected to an iterative categorization process, through which the extracted elements were grouped into conceptual dimensions capable of spanning all models in a non-redundant manner, following the procedure proposed for the construction of conceptual frameworks \cite{jabareen2009building}. The dimensions thus identified served as the basis for the structural comparison of the models, operationalized through a model~$\times$~dimension analytical matrix inspired by the \textit{concept matrix} proposed by \autorano{webster2002analyzing}. Each cell of the matrix was populated based on passages identified in the primary sources and, when pertinent, in the critical review studies included in the corpus, ensuring traceability between the recorded attributes and their bibliographic origin and avoiding extrapolations to contexts not supported by the analyzed sources \cite{hong2014framework, webster2002analyzing}.

From the patterns observed in the comparative matrix, operational decision criteria were derived, associating characteristics identified in the models with specific research conditions. Each comparative dimension generated one or more criteria that make explicit under which conditions of context, type of use, adoption stage, or level of analytical complexity a given model tends to be more suitable \cite{whetten1989constitutes, jabareen2009building, mkhomazi2013guide}. The criteria were formulated so as to be usable regardless of the researcher's prior knowledge of each model, making the selection process less dependent on familiarity and more dependent on the correspondence between the model's characteristics and the research problem under investigation \cite{turner2010does, hong2014framework}.

\begin{figure}[ht]
\centering
\pdftooltip{\includegraphics[width=.6\textwidth,trim=0 5cm 0 5cm,clip]{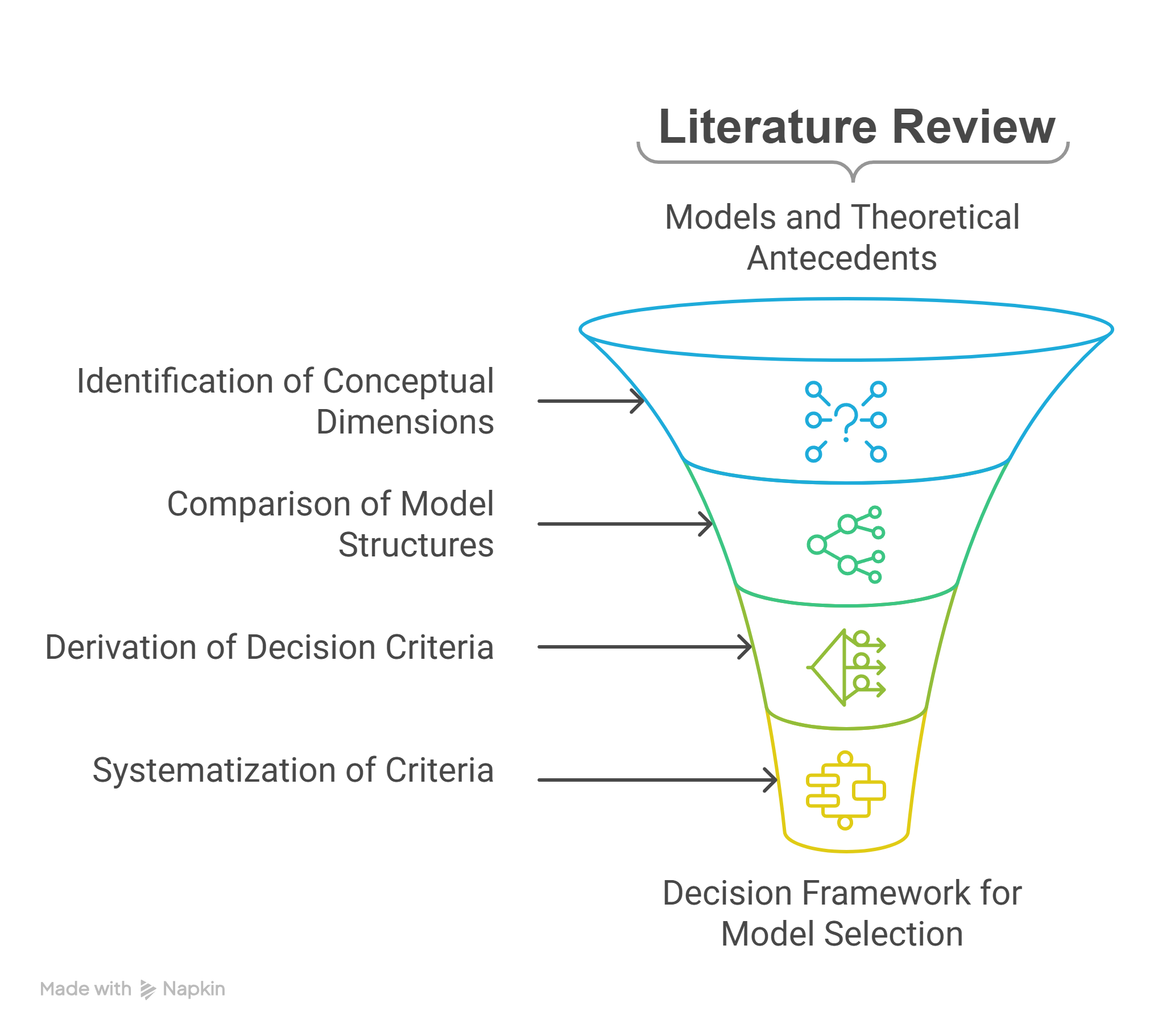}}{The figure illustrates the study's methodological approach using a funnel diagram to represent the process of filtering and refining information. At the top, above the wide opening of the funnel, is the heading Literature Review, followed by the subtitle Theoretical Models and Antecedents. The process is detailed through four stages described on the left, with arrows pointing to the corresponding sections of the funnel. The first stage is the Identification of Conceptual Dimensions. The second stage is the Comparison of Model Structures. The third stage consists of the Derivation of Decision Criteria. The fourth and final stage is the Systematization of Criteria. At the bottom, the final result is the Decision Framework for Model Selection.}
\caption{Overview of the approach adopted for deriving the decision framework for the selection of technology acceptance and use models.}
\label{fig:abordagemEstudo}
\end{figure}

Finally, the derived criteria were systematized into a decision framework for the selection of technology acceptance and use models, conceived as a research decision-support artifact \cite{hevner2004design}. Figure~\ref{fig:abordagemEstudo} synthesizes this sequence, showing the trajectory that starts from the literature review and the examination of models and theoretical antecedents, proceeds through the identification of conceptual dimensions, the comparison of model structures, and the derivation of decision criteria, culminating in the systematization of the criteria into the proposed framework.

\FloatBarrier

\section{Comparative Analysis of the Models} \label{sec:AnaliseComparativa}

Table~\ref{qua:comparacao_modelos} summarizes the structural comparison of the five models according to the conceptual dimensions identified in Section~\ref{sec:AbordagemEstudo}. The following paragraphs develop in prose the most relevant convergences, divergences, and analytical implications for model selection in research.\par
\antesquadro

{\scriptsize
\setlength{\tabcolsep}{2pt}
\captionsetup{type=table,name=Table}
\begin{longtable}{|>{\raggedright\arraybackslash}p{2.60cm}|>{\raggedright\arraybackslash}p{2.29cm}|>{\raggedright\arraybackslash}p{2.29cm}|>{\raggedright\arraybackslash}p{2.29cm}|>{\raggedright\arraybackslash}p{2.29cm}|>{\raggedright\arraybackslash}p{2.29cm}|}
\caption{Conceptual comparison of technology acceptance and use models.}\label{qua:comparacao_modelos}\\
\hline
\textbf{Comparative dimension} & \textbf{\ac{TAM}} & \textbf{\ac{TAM2}} & \textbf{\ac{TAM3}} & \textbf{\ac{UTAUT}} & \textbf{\ac{UTAUT2}} \\
\hline
\endfirsthead

\endhead

\hline
\multicolumn{6}{r}{\scriptsize\textit{Continued on the next page}}\\
\endfoot

\hline
\endlastfoot

\textbf{Predominant explanatory focus}
& Behavioral intention based on perceived usefulness and perceived ease of use.
& Behavioral intention with expansion of perceived usefulness through social influence and instrumental processes.
& Perceived ease of use and its individual and contextual antecedents.
& Intention and use based on performance expectancy, effort expectancy, social influence, and facilitating conditions.
& Intention and use in a consumer context, incorporating hedonic motivation, price value, and habit. \\
\hline

\textbf{Nature of determinants}
& Predominantly cognitive and individual.
& Cognitive, social, and job-related.
& Cognitive, individual, contextual, and organizational.
& Cognitive, social, contextual, and moderated by individual characteristics.
& Cognitive, social, contextual, and experiential. \\
\hline

\textbf{Social influence}
& Not made explicit as a core construct.
& Explicitly incorporated through subjective norm and image.
& Maintained from the structure inherited from \ac{TAM2}.
& Core construct of the model.
& Construct maintained, now situated within a predominantly voluntary consumer context. \\
\hline

\textbf{Intended type of use}
& Voluntary or mandatory, with no explicit treatment of this distinction in the original formulation.
& Explicitly distinguishes voluntary and mandatory use, with voluntariness operating as a moderator of social influence.
& Voluntary or mandatory, with less emphasis on the distinction between use regimes than on the influence of experience on constructs.
& Voluntary and mandatory, with voluntariness of use as a moderator of the relationships among constructs.
& Predominantly voluntary, oriented toward individual consumption and without organizational imposition. \\
\hline

\textbf{Moderators}
& Not emphasized in the original formulation.
& Experience and voluntariness of use.
& Experience, particularly as a variable that alters the strength of relationships between antecedents and use perceptions.
& Gender, age, experience, and voluntariness of use.
& Gender, age, and experience. \\
\hline

\textbf{Adoption stage considered}
& Initial acceptance, with focus on the formation of behavioral intention.
& Initial acceptance, with emphasis on social influences and use conditions in the organizational context.
& Acceptance and evolution of use, with attention to variations arising from experience.
& Acceptance and use, considered under the effect of individual and contextual moderators.
& Acceptance, use, and continuance, with incorporation of habit in consolidated behavior. \\
\hline

\textbf{Primary application context}
& Technology adoption in diverse contexts, with strong use in organizational settings.
& Organizational and work contexts.
& Organizational contexts requiring greater detail regarding use barriers.
& Organizational contexts, particularly associated with job performance.
& Individual consumer contexts and voluntary use. \\
\hline

\textbf{Structural complexity}
& Low, with a parsimonious structure.
& Moderate, with controlled expansion of the original model.
& High, due to the detailed treatment of antecedents and relationships.
& High, by integrating multiple constructs and moderators.
& High, with expansion of \ac{UTAUT} through the incorporation of constructs specific to the consumer context. \\
\hline

\textbf{Analytical implication for selection}
& Suitable when the interest lies in core beliefs and a lean explanatory structure.
& Suitable when adoption involves social influence and task characteristics in an organizational context.
& Suitable when greater analytical granularity regarding ease of use and interaction conditions is sought.
& Suitable when the study requires an integrative perspective with individual heterogeneity and organizational context.
& Suitable when the phenomenon under investigation involves consumption, voluntariness, experience, and habit. \\

\end{longtable}
\fontequadro{Prepared by the authors, based on \autorano{davis1989perceived}; \autorano{venkateshTheoreticalExtensionTechnology2000b}; \autorano{venkateshUserAcceptanceInformation2003b}; \autorano{venkateshTechnologyAcceptanceModel2008}; \autorano{venkateshConsumerAcceptanceUse2012a}.}
}

A first relevant distinction among the models concerns the explanatory focus that guides each formulation and, consequently, the type of determinant that assumes centrality in its structure. In \ac{TAM}, the explanation of behavioral intention centers on individual cognitive beliefs, particularly perceived usefulness and perceived ease of use, which gives the model a more parsimonious character and a predominantly instrumental orientation \cite{davis1989perceived}. \ac{TAM2} preserves this foundation but expands it by incorporating social influence mechanisms and cognitive instrumental processes associated with the job \cite{venkateshTheoreticalExtensionTechnology2000b}, making the influence of the organizational environment on the formation of perceived usefulness more visible. \ac{TAM3}, in turn, deepens the analysis of the antecedents of perceived ease of use, increasing the granularity of the explanation by including individual, contextual, and organizational factors that influence the experience of interacting with the technology \cite{venkateshTechnologyAcceptanceModel2008}.

This trajectory of expansion becomes more evident with \ac{UTAUT}, which abandons the logic of incremental extension of a single model and operates instead through theoretical integration \cite{venkateshUserAcceptanceInformation2003b}. Its structure brings together different explanatory traditions in a more comprehensive arrangement, in which performance expectancy, effort expectancy, social influence, and facilitating conditions are articulated with moderators associated with individual and situational heterogeneity. \ac{UTAUT2} preserves this integrative logic but shifts it to the consumer context, incorporating hedonic motivation, price value, and habit \cite{venkateshConsumerAcceptanceUse2012a}. Accordingly, the comparison reveals an important conceptual movement: while the \ac{TAM} lineage remains strongly anchored in individual perceptions of usefulness and effort, \ac{UTAUT} and, above all, \ac{UTAUT2} accommodate social, contextual, and experiential elements more explicitly, broadening the scope of explanation of use behavior.

Another central comparative dimension concerns the intended type of use and the nature of the relationship between the user and the technology presupposed by each model. In \ac{TAM}, this relationship is treated in a relatively general manner, without an explicit distinction between voluntary and mandatory use regimes. \ac{TAM2} makes this difference analytically more visible by incorporating voluntariness as a moderator of social influence \cite{venkateshTheoreticalExtensionTechnology2000b}, which makes it particularly sensitive to organizational scenarios in which adoption may be partly induced by norms, expectations, and institutional requirements. \ac{TAM3} maintains this background but focuses less on the distinction between use regimes and more on the variation of antecedents as a function of the user's accumulated experience. \ac{UTAUT}, for its part, makes explicit the coexistence of voluntary and mandatory use and treats this difference as part of the model's moderating structure \cite{venkateshUserAcceptanceInformation2003b}. In \ac{UTAUT2}, by contrast, use is predominantly conceived as voluntary, driven by consumer experience and less dependent on organizational imposition \cite{venkateshConsumerAcceptanceUse2012a}. This difference is not merely contextual but conceptual, as it alters the type of rationality that underpins technology adoption.

The comparison also reveals distinctions regarding the adoption stage emphasized by each model. \ac{TAM} and \ac{TAM2} remain closer to initial acceptance, with emphasis on the formation of behavioral intention and the factors that influence it. \ac{TAM3} partially shifts this focus by considering the effects of experience on the stability or change of user perceptions over time \cite{venkateshTechnologyAcceptanceModel2008}. \ac{UTAUT}, while maintaining the explanation of intention as a central component, articulates it more directly with use behavior, considering moderators that operate in the transition between acceptance and actual use. \ac{UTAUT2} deepens this movement by incorporating habit, allowing the treatment not only of intention or initial adoption but also of continuance of use in contexts where the technology has already been incorporated into the user's practices \cite{venkateshConsumerAcceptanceUse2012a}. This difference is methodologically relevant because it indicates that the models differ not only in the type of construct they emphasize but also in the adoption stage they prioritize.

Important differences are also observed in the primary application context. \ac{TAM} and its extensions have been widely employed in organizational studies, although their use has expanded to other settings. In \ac{TAM2}, this link is particularly clear, as the model explicitly incorporates job relevance, image, and subjective norm, elements strongly associated with the work environment \cite{venkateshTheoreticalExtensionTechnology2000b}. \ac{TAM3} maintains this foundation and may be especially useful when the interest lies in barriers, support, and conditions that affect ease of use in more complex interaction situations \cite{mohammadzadeh2025identifying}. \ac{UTAUT} also remains strongly associated with organizational contexts, now under a broader formulation oriented toward performance, infrastructure, and individual heterogeneity. \ac{UTAUT2}, on the other hand, represents a clear conceptual shift by relocating the center of explanation to use in individual consumer contexts \cite{venkateshConsumerAcceptanceUse2012a}. This shift alters not only the empirical setting of application but the very type of user-technology relationship the model seeks to explain \cite{hong2014framework}.

At the structural level, the models also differ in the degree of complexity and the analytical granularity they require. \ac{TAM} is the most parsimonious in the lineage, favoring investigations that require lean explanatory structures and a focus on core determinants \cite{davis1989perceived, turner2010does}. \ac{TAM2} expands this structure without breaking with its basic logic, maintaining relative conceptual economy. \ac{TAM3} increases the structural complexity of the model more significantly by detailing antecedents and relationships previously treated in a more aggregated manner \cite{venkateshTechnologyAcceptanceModel2008}. \ac{UTAUT} and \ac{UTAUT2} also exhibit greater structural breadth, combining multiple determinants and moderators in a formulation more sensitive to the heterogeneity of contexts and users \cite{williams2015unified, debritoLimitacoesDosModelos2019a, dwivedi2019re, batista2023consideraccoes}. In analytical terms, this means that the choice among models involves not only deciding which constructs are most relevant but also which level of detail and complexity is compatible with the research problem and the type of explanation intended.

Taken together, these differences should not be interpreted in terms of the superiority of one model over another, but as indicators of suitability for different analytical objectives. More parsimonious models tend to be more compatible with investigations that prioritize core beliefs and leaner explanatory structures, whereas more comprehensive models offer greater sensitivity to individual heterogeneity, contextual variations, and more complex use dynamics \cite{debritoLimitacoesDosModelos2019a, malatji2020understanding, batista2023consideraccoes, royApplyingUTAUT2Qualitative2025a}. Thus, the comparison shows that theoretical selection depends less on the frequency with which a given model is used in the literature and more on the correspondence between its conceptual characteristics and the phenomenon under investigation \cite{mkhomazi2013guide, hong2014framework}. These differences provide the basis for deriving operational criteria to guide the researcher in choosing the most suitable model for their context and objectives, as developed in the following section.

\section{Decision Framework for Model Selection}\label{sec:Framework}

\begin{figure}[ht]
\centering
\pdftooltip{\includegraphics[width=1\textwidth]{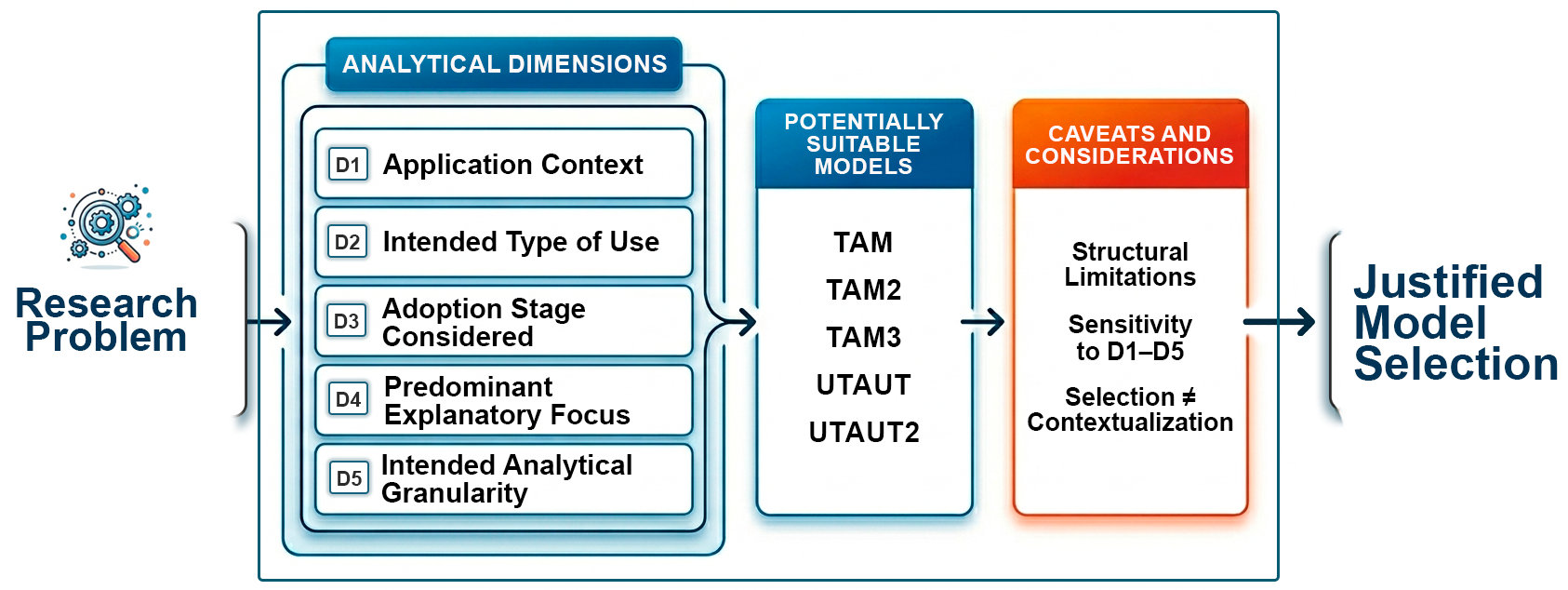}}{The figure presents the overall architecture of the decision framework for selecting technology acceptance and use models. On the left, the entry point is the Research Problem. The flow moves to a block labeled Analytical Dimensions, containing five numbered dimensions: D1 Application Context, D2 Intended Type of Use, D3 Adoption Stage Considered, D4 Predominant Explanatory Focus, and D5 Intended Analytical Granularity. The output of this block feeds into a second block labeled Potentially Suitable Models, listing TAM, TAM2, TAM3, UTAUT, and UTAUT2. This block connects to a third block labeled Caveats and Considerations, containing three items: Structural Limitations, Sensitivity to D1 through D5, and Selection is not the same as Contextualization. The final output on the right is the Justified Model Selection.}
\caption{Overview of the decision framework for the selection of technology acceptance and use models.}
\label{fig:frameworkModelo}
\end{figure}

This section presents the proposed decision framework for guiding the structured selection of technology acceptance and use models. The proposal articulates four components: a visual representation of the framework architecture, a table of guiding questions that operationalizes the analytical dimensions, a table of model suitability profiles, and a set of caveats and considerations that delimit the conditions for using the proposal. Figure~\ref{fig:frameworkModelo} presents the overall view of the framework, showing the trajectory that starts from the characterization of the research problem, proceeds through the five analytical dimensions (D1 to D5), leads to the identification of potentially suitable models, and then subjects the recommendation to a set of caveats before a justified selection is made.

\FloatBarrier

The trajectory represented in Figure~\ref{fig:frameworkModelo} uses the research problem formulated by the researcher as its entry point, typically related to understanding how and why individuals accept, adopt, or continue to use a given technology in specific contexts. The formulation of the problem does not require all study conditions to be predefined, but it does require the researcher to be clear about what they intend to investigate. The five analytical dimensions (D1 to D5) organize the reflection on this problem through guiding questions that lead the researcher to make explicit aspects that may be implicit or still undefined, such as the context in which the technology will be used, the expected conditions of use, and the nature of the intended explanation. This characterization constitutes the basis from which the framework guides model selection, as detailed in the tables presented below.

Of the nine comparative dimensions presented in Table~\ref{qua:comparacao_modelos}, four were directly incorporated as framework dimensions: primary application context (D1), intended type of use (D2), adoption stage considered (D3), and structural complexity, renamed as intended analytical granularity (D5) to reflect the researcher's perspective rather than a fixed property of the model. Three conceptually related dimensions (predominant explanatory focus, nature of determinants, and social influence) were merged into a single dimension (D4), operationalized through one main guiding question and three refinement questions, detailed in the supplementary material, that capture, respectively, the nature of the prioritized determinants, the role of social influence, and the relevance of affective, hedonic, or habit-related elements. The two remaining dimensions (moderators and analytical implication for selection) are not part of the framework dimensions: moderators operate as internal variables of the models and belong to the post-selection operationalization stage, while analytical implication constitutes a consequence derived from the other dimensions, whose function is now fulfilled by the suitability profiles.

It should be noted that the profile of \ac{TAM3} was derived primarily from its original formulation \cite{venkateshTechnologyAcceptanceModel2008} and from broader comparative analyses, as the critical review corpus does not include a study dedicated exclusively to this model, which is acknowledged as a limitation in Section~\ref{sec:DiscussaoLimitacoes}.

The five dimensions operate in an analytical sequence, from broader delimitation to finer refinement. Application context (D1) and intended type of use (D2) function as primary filters, since the distinction between organizational and individual consumer settings, as well as between voluntary and mandatory use regimes, substantially reduces the set of candidate models.

The adoption stage considered (D3) refines the selection by distinguishing studies focused on initial acceptance from those investigating continued use or multiple stages. The predominant explanatory focus (D4) assumes a central role when the preceding filters are not sufficient to differentiate models that share similar context and stage, such as \ac{TAM3}, \ac{UTAUT}, and \ac{UTAUT2}. The intended analytical granularity (D5) complements the analysis by making explicit the tension between conceptual economy and explanatory coverage, although its discriminatory power is more limited among models of high structural complexity.

To operationalize these dimensions, the framework provides a set of 16 guiding questions, distributed as follows: three for D1, three for D2, three for D3, four for D4, and three for D5, which together lead the researcher to position their study within each of the five analytical dimensions. The questions do not introduce additional complexity to the selection process but rather make explicit aspects that typically underlie the reflection involved in theoretical choice. Table~\ref{qua:indagacoes_orientadoras} summarizes the main guiding question of each dimension, identified as D1.1 to D5.1, together with response alternatives and refinement guidelines for cases requiring further analysis.

\antesquadro

{\scriptsize
\setlength{\tabcolsep}{3pt}
\captionsetup{type=table,name=Table}
\begin{longtable}{|>{\raggedright\arraybackslash}p{2.2cm}|>{\raggedright\arraybackslash}p{8.6cm}|>{\raggedright\arraybackslash}p{3.6cm}|}
\caption{Main guiding questions of the decision framework.}\label{qua:indagacoes_orientadoras}\\
\hline
\textbf{Dimension} & \textbf{Guiding question} & \textbf{Distinction and refinement} \\
\hline
\endfirsthead

\endhead

\hline
\multicolumn{3}{r}{\scriptsize\textit{Continued on the next page}}\\
\endfoot

\hline
\endlastfoot

\textbf{D1. Application Context}
& \textbf{D1.1: Primary classification of context:} In which predominant setting will the investigated technology be used? (a) Organizational, with use associated with tasks, routines, or institutional responsibilities. (b) Individual consumption, with use associated with personal choices of the user. (c) Hybrid, with relevant elements from both contexts.
& Distinguishes organizational, individual consumption, and hybrid contexts. In hybrid cases, consider the role of the user and the predominant motivation for use. \\
\hline

\textbf{D2. Intended Type of Use}
& \textbf{D2.1: Primary classification of the use regime:} What is the use regime presupposed by the context under investigation? (a) Voluntary, when the adoption and use of the technology depend predominantly on the user's choice. (b) Mandatory, when there is formal or institutional imposition for the use of the technology. (c) Mixed, when voluntary and mandatory elements coexist in an analytically relevant manner.
& Distinguishes models sensitive to use regime and voluntariness. In mixed cases, consider the difference between formal and perceived voluntariness. \\
\hline

\textbf{D3. Adoption Stage Considered}
& \textbf{D3.1:  Primary classification of the adoption stage:} Which stage of the adoption process does the research predominantly investigate? (a) Initial acceptance, with focus on the formation of intention and the initial decision to adopt the technology. (b) Continued use, with focus on use behavior after the initial adoption of the technology. (c) Multiple stages, when initial acceptance and continued use are both central to the analytical objectives of the research.
& Distinguishes models oriented toward initial intention and models sensitive to consolidated use. In multi-stage studies, identify which stage carries greater weight in the analytical objectives. \\
\hline

\textbf{D4. Predominant Explanatory Focus}
& \textbf{D4.1: Predominant nature of the determinant:} What type of determinant does the study prioritize to explain technology acceptance or use? (a) Predominantly cognitive-instrumental, with focus on beliefs about usefulness and effort associated with use. (b) Cognitive-instrumental expanded by social, contextual, or use-experience-related factors. (c) Multiple types of determinants, including cognitive, social, contextual, and affective factors, articulated in an integrated structure.
& Distinguishes cognitive-instrumental models and models with a broader explanatory structure. When necessary, distinguish the role of social influence and the relevance of affective, hedonic, or habit-related elements. \\
\hline

\textbf{D5. Intended Analytical Granularity}
& \textbf{D5.1: Primary classification of intended granularity:} What level of analytical granularity does the study intend to achieve in explaining the phenomenon? (a) Low granularity, with a lean explanatory structure focused on core determinants. (b) Intermediate granularity, with controlled expansion of the explanatory structure to capture additional relevant determinants. (c) High granularity, with a detailed explanatory structure to incorporate multiple antecedents, moderators, and relationships among constructs.
& Distinguishes more parsimonious models from more complex and comprehensive ones. In intermediate situations, consider the tension between conceptual economy and explanatory coverage. \\
\hline

\end{longtable}
\fontequadro{Prepared by the authors. The expanded set of guiding questions detailed by dimension is available as supplementary material \cite{anonymous_2026_19714800}.}
}

\FloatBarrier

The main question of D4 (D4., Table~\ref{qua:indagacoes_orientadoras}) is complemented by three refinement questions,
detailed in the supplementary material \cite{anonymous_2026_19714800}. The first (D4.2) asks about the role of social influence in the investigation, distinguishing studies in which it is not central, studies requiring normative or status-related social influence, and studies requiring a broader formulation as expectations of relevant others. The second (D4.3) asks whether the study requires sensitivity to affective, hedonic, or habit-related elements, distinguishing instrumental-only focuses from those requiring attention to enjoyment, anxiety, or consumer-oriented constructs such as hedonic motivation, habit, and price value. The third (D4.4) provides an internal tie-breaking rule when D4.2 and D4.3 point to different models, asking which aspect carries the greatest weight in the explanatory objectives of the research.

Table~\ref{qua:perfis_modelos} summarizes the suitability profiles of the five models according to the framework dimensions. Each profile describes the typical application conditions of the model, organized by D1 to D5, based on the comparative analysis presented in Section~\ref{sec:AnaliseComparativa}.

\antesquadro

{\scriptsize
\setlength{\tabcolsep}{2pt}
\captionsetup{type=table,name=Table}
\begin{longtable}{|>{\raggedright\arraybackslash}p{1.35cm}|>{\raggedright\arraybackslash}p{2.95cm}|>{\raggedright\arraybackslash}p{2.30cm}|>{\raggedright\arraybackslash}p{2.30cm}|>{\raggedright\arraybackslash}p{3.55cm}|>{\raggedright\arraybackslash}p{1.65cm}|}
\caption{Suitability profiles of technology acceptance and use models according to the framework dimensions.}\label{qua:perfis_modelos}\\
\hline
\textbf{Model} & \textbf{Typical context (D1)} & \textbf{Use regime (D2)} & \textbf{Primary stage (D3)} & \textbf{Explanatory core (D4)} & \textbf{Granularity level (D5)} \\
\hline
\endfirsthead

\endhead

\hline
\multicolumn{6}{r}{\scriptsize\textit{Continued on the next page}}\\
\endfoot

\hline
\endlastfoot

\textbf{\ac{TAM}}
& Contexts focused on the individual relationship with the technology, with strong use in organizational settings.
& Does not explicitly distinguish voluntary from mandatory use.
& Initial acceptance.
& Perceived usefulness and perceived ease of use.
& Low \\
\hline

\textbf{\ac{TAM2}}
& Organizational and work contexts.
& Sensitive to the distinction between voluntary and mandatory use.
& Initial acceptance.
& \ac{TAM} expanded by social influence and job-related factors.
& Intermediate \\
\hline

\textbf{\ac{TAM3}}
& Organizational contexts requiring detailed treatment of use barriers.
& Less emphasis on use regime than on experience.
& Acceptance and variation of perceptions with experience.
& Antecedents of perceived ease of use (anchor and adjustment).
& High \\
\hline

\textbf{\ac{UTAUT}}
& Performance-oriented organizational contexts.
& Explicitly addresses voluntary and mandatory use.
& Acceptance and use.
& Performance expectancy, effort expectancy, social influence, and facilitating conditions.
& High \\
\hline

\textbf{\ac{UTAUT2}}
& Individual consumer contexts.
& Predominantly voluntary.
& Acceptance, use, and continuance.
& \ac{UTAUT} expanded by hedonic motivation, price value, and habit.
& High \\
\hline
\end{longtable}
\fontequadro{Prepared by the authors, based on the comparative analysis of the models (Section~\ref{sec:AnaliseComparativa}).}
}

\FloatBarrier

The combined use of the two tables constitutes the central mechanism of the framework. The responses to the guiding questions (Table~\ref{qua:indagacoes_orientadoras}) produce a characterization of the study that, when compared against the suitability profiles (Table~\ref{qua:perfis_modelos}), allows the identification of the model whose conceptual structure most closely matches the conditions stated by the researcher.

Before consolidating the selection, the Caveats and Considerations block (Figure~\ref{fig:frameworkModelo}) brings together three considerations that must be addressed for model selection to be effectively justified. The first concerns structural limitations. The framework operates on five specific models of the technology acceptance and use lineage, and the comparison between the study characterization and the profiles may yield three types of outcome: clear correspondence with a single profile, partial correspondence with two or more profiles, or absence of adequate correspondence with any of the presented profiles. The latter two cases do not represent a failure in the use of the framework, but rather reflect properties of the analyzed model space and of the nature of the research problem. When correspondence is partial, the researcher should consider which profile best approximates their characterization, using dimensions D1 to D5 as tie-breaking criteria. When no profile is adequate, the research problem may exceed the scope of the models covered. Additionally, the dimensions do not have uniform discriminatory power across all models. Dimension D5, for example, differentiates models of low and intermediate granularity (\ac{TAM} and \ac{TAM2}), but does not discriminate among models of high granularity (\ac{TAM3}, \ac{UTAUT}, and \ac{UTAUT2}), whose differentiation depends predominantly on D4.

The second consideration concerns sensitivity to dimensions D1 to D5. The recommendation produced by the framework depends directly on the characterization made by the researcher across the five dimensions. If, over the course of the investigation, any of these conditions is reinterpreted (for example, if the context initially classified as organizational reveals significant elements of individual consumption, or if the use regime initially considered voluntary reveals unforeseen mandatory components), the correspondence between the characterization and the model profiles may change. In such cases, the researcher should revisit the dimensions and verify whether the selected model remains aligned with the new configuration of the problem. The framework does not operate as an irreversible decision, but as guidance dependent on the conditions stated at the time of application.

The third consideration concerns the distinction between selection and contextualization. The framework guides the selection of the theoretical model, not its adaptation to the specific research context. Choosing a model from among the five analyzed constitutes a prior and distinct decision from the subsequent contextualization, which involves adjustments to constructs, causal relationships, or moderators to suit the phenomenon under investigation. This distinction is consistent with the perspective that selection and contextualization operate as sequential and complementary stages of the theoretical grounding process in empirical research \cite{hong2014framework}.

Once these caveats have been considered, the researcher has the necessary elements for a justified model selection, as represented in the final stage of Figure~\ref{fig:frameworkModelo}. The selection justification is built from the stated correspondence between the characterization of the research problem across dimensions D1 to D5 and the suitability profile of the chosen model, making the decision process explicit, traceable, and less dependent on the researcher's prior familiarity with the models of the lineage. In this way, the framework addresses the research question (\ac{RQ}) of this study by offering a structured and conceptually grounded path for the selection of technology acceptance and use models, considering different research contexts and objectives. The following section demonstrates the application of the framework to contrasting research scenarios.

\section{Application of the Decision Framework for Model Selection} \label{sec:AplicacaoFrameworkDecisaoSelecao}

The studies used in this demonstration were selected based on five criteria: (a) the study adopts one of the five models as a theoretical basis for investigating acceptance, use, or behavioral intention regarding a technology; (b) the contexts are contrasting along dimensions D1 and D2 of the framework, allowing different profiles to be examined; (c) the declared theoretical model is identifiable and is not embedded in hybrid formulations that prevent comparison with the original model structure; (d) the studies are available in open access, enabling independent verification of the demonstration; and (e) among the studies that met the preceding criteria, preference was given to works published in Brazilian computing venues, resorting to international venues when coverage of the models or contexts required it.

Two studies were selected: the first investigates the acceptance of a management system in an organizational context with institutional use \cite{pinheiroAvaliacaoUsabilidadeSistema2023a}, and the second examines the intention to adopt wearable devices among older consumers in a voluntary use context \cite{wu2024investigating}. The framework is applicable to studies that adopt the analyzed models as a theoretical basis for investigating acceptance, use, or behavioral intention under conditions aligned with the proposed dimensions, and the demonstration that follows is illustrative, not exhaustive.

The study by \autorano{pinheiroAvaliacaoUsabilidadeSistema2023a} evaluates the usability of \ac{SIGProj}, currently used at a Brazilian federal public university, and investigates the acceptance of a proposed new system for university extension activities. \ac{SIGProj} is provided by the Ministry of Education to federal universities for the registration and management of extension projects, constituting an institutional system whose use is associated with the performance of employees' formal duties. To assess the acceptance of the proposed system, the authors conducted a workshop with fourteen expert users and administered a questionnaire based on constructs of perceived usefulness, perceived ease of use, satisfaction, and behavioral intention, declaring the adoption of \ac{TAM3} as the reference model.

The application of the guiding questions (Table~\ref{qua:indagacoes_orientadoras}) to this scenario proceeds through the five framework dimensions. In D1, the system is institutional and the users are employees performing formal duties, which characterizes an organizational context. In D2, \ac{SIGProj} is the system designated by the Ministry of Education for managing extension activities at federal universities. Assuming that the proposed replacement would operate under the same institutional-use regime, this places the scenario closer to a use regime with limited voluntariness. In D3, the object of evaluation is a proposed system, not yet implemented, which situates the investigation at the initial acceptance stage. In D4, the operationalized constructs do not include social influence, hedonic motivation, or affective antecedents of ease of use, indicating a predominantly cognitive-instrumental focus. In D5, the instrument does not detail antecedents of perceived usefulness or perceived ease of use, does not operationalize moderators, and does not employ the anchor and adjustment logic characteristic of \ac{TAM3}, which indicates low granularity.

This characterization reveals a tension between the label declared by the authors and the actual operationalization of the study. \ac{TAM3} presupposes the modeling of antecedents of perceived ease of use organized under the anchor and adjustment logic, such as computer self-efficacy, computer anxiety, perceptions of external control, and objective usability, among others \cite{venkateshTechnologyAcceptanceModel2008}. None of these antecedents is operationalized in the administered instrument. The constructs actually employed, perceived usefulness, perceived ease of use, and behavioral intention, correspond to the core structure of \ac{TAM} \cite{davis1989perceived}. Satisfaction, present in the instrument, is not part of the original model formulation and constitutes an addition by the authors to the evaluation design.

When comparing the responses obtained across the five dimensions against the model suitability profiles (Table~\ref{qua:perfis_modelos}), the framework indicates correspondence with the profile of \ac{TAM}: organizational context, cognitive-instrumental focus, low granularity, and initial acceptance. Additionally, the condition of limited voluntariness identified in D2 suggests that incorporating social influence constructs could broaden the explanatory power of the investigation, bringing the scenario closer to the profile of \ac{TAM2} \cite{venkateshTheoreticalExtensionTechnology2000b}. The framework, therefore, does not invalidate the authors' choice, but identifies that the operationalization adopted is more compatible with \ac{TAM} than with the declared \ac{TAM3}, and signals that the institutional context could justify considering \ac{TAM2} as an alternative.

The study by \autorano{wu2024investigating}, published in the journal Frontiers in Public Health, investigates the factors that influence the willingness of older consumers to adopt smart wearable devices for health monitoring in China. The theoretical model adopted is \ac{UTAUT2}, whose specification in this study preserves six of the seven constructs of the original model (performance expectancy, effort expectancy, social influence, facilitating conditions, hedonic motivation, and price value). The authors exclude the habit construct based on the participants' lack of prior experience with the investigated technology, integrate \ac{TRI} as an antecedent of performance expectancy and effort expectancy, incorporate digital health literacy as a moderating variable, and omit the standard moderators of \ac{UTAUT2} (gender, age, and experience). Data collection involved a questionnaire administered to 389 older adult respondents, and the analysis employed structural equation modeling.

The application of the guiding questions (Table~\ref{qua:indagacoes_orientadoras}) to this scenario yields a characterization that contrasts with the previous case. In D1, the investigated technology is a consumer product acquired individually (commercial wearable devices), and the users are end consumers with no organizational obligation associated with use, which characterizes an individual consumer context. In D2, the decision to adopt the device is entirely personal, with no institutional imposition, indicating voluntary use. In D3, the participants have no prior experience with the technology, and the study measures behavioral intention, which situates the investigation at the initial acceptance stage.

In D4, the model operationalizes performance expectancy, effort expectancy, social influence, facilitating conditions, hedonic motivation, and price value, articulating cognitive, social, and affective determinants in an integrated structure. The explanatory focus is therefore one of theoretical integration, with sensitivity to hedonic motivations and cost-benefit evaluation. In D5, the study employs six constructs from \ac{UTAUT2}, four dimensions of \ac{TRI} as antecedents, and digital health literacy as a moderator, indicating high granularity.

When comparing this characterization against the model suitability profiles (Table~\ref{qua:perfis_modelos}), the framework indicates correspondence with the profile of \ac{UTAUT2}: individual consumer context, voluntary use, theoretical integration with hedonic motivation and price value, and high granularity. The authors' choice coincides with the framework recommendation. The exclusion of the habit construct is consistent with the participants' lack of prior experience with the investigated technology, a condition under which habit is not expected to operate as a determinant. This condition aligns with the characterization obtained in D3, which situates the investigation at the initial acceptance stage. The framework thus makes it possible to recognize that this adaptation does not alter the essential character of the model, but reflects a methodological decision aligned with the adoption stage under investigation.

The integration of \ac{TRI} as an antecedent of performance expectancy and effort expectancy constitutes a contextualization decision subsequent to the selection of the base model, consistent with the distinction proposed in Section~\ref{sec:Framework} between theoretical selection and contextualization \cite{hong2014framework}. The framework guides the first stage, and the incorporation of technology readiness variables belongs to the second. In this sense, the case illustrates the complementarity between the two stages: \ac{UTAUT2} is selected based on conceptual correspondence with the research problem, and \ac{TRI} is integrated as a contextualization resource for the investigated population.

The two scenarios examined illustrate distinct modes of operation of the framework. In the first, the artifact identified a divergence between the declared model and the actual operationalization, suggesting alternatives more consistent with the research design. In the second, the artifact confirmed the authors' choice and made it possible to recognize that the adaptation carried out was compatible with the study conditions. In both cases, the trajectory through the dimensions and the comparison against the suitability profiles produced traceable characterizations, less dependent on prior familiarity with the models, demonstrating the framework's capacity to structure and ground the theoretical selection decision.

\section{Discussion and Limitations}\label{sec:DiscussaoLimitacoes}

The central contribution of the framework lies in structuring a decision that, in the literature, remains predominantly implicit. Unlike studies that compare models by explanatory power or empirical performance \cite{brown2015technology, taherdoost2018review}, the proposed artifact guides the theoretical selection process based on conceptual criteria derived from the structural differences among the models. This guidance is more specific than that offered by generic theoretical selection guides \cite{mkhomazi2013guide}, as it operates on a delimited set of models with dimensions extracted from their original formulations. It is also positioned upstream of the contribution of \autorano{hong2014framework}, who guides the contextualization of models already chosen: the proposed framework organizes the stage that precedes contextualization, making the two proposals sequential and complementary. For the Brazilian \ac{HCI} community, in which critical reviews dedicated to these models were not identified in an exploratory search of the main venues of the field, the artifact may serve as a structured reference for theoretical grounding in technology acceptance and use research.

The demonstration illustrated that the framework operates both through constructive divergence and through conceptually grounded confirmation, reinforcing its role as an instrument of structured reflection, not as a rigid prescription. From a methodological standpoint, this dual capacity produces gains that go beyond selection itself: it makes the theoretical justification verifiable by third parties, exposes misalignments between the declared model and the actual operationalization, and offers explicit criteria for recognizing when adaptations are consistent with the study conditions. The organizational scenario exemplified the exposure of misalignment by revealing that the operationalization was more compatible with \ac{TAM} than with the declared \ac{TAM3}. The consumer scenario exemplified the recognition of adaptations by showing that the exclusion of habit did not alter the essential character of \ac{UTAUT2}.

These contributions should be read in light of the artifact's limitations. The evaluation was conducted through an illustrative demonstration applied to two contrasting scenarios, consistent with \ac{DSR} principles \cite{hevner2004design}, but without empirical validation with researchers engaged in an actual theoretical selection process. The critical review corpus does not include a study dedicated exclusively to \ac{TAM3}, which may affect the robustness of the corresponding profile. The scope is restricted to five models of the \ac{TAM}/\ac{UTAUT} lineage, not encompassing models from other theoretical traditions. The application of the framework involves interpretive judgment on the part of the researcher, especially in hybrid or borderline scenarios, which may lead to different classifications in the absence of additional calibration protocols. As a trade-off for its operational character, the artifact translates models with dense conceptual structures into synthetic five-dimension profiles, which facilitates decision-making but does not replace an in-depth reading of the original formulations.

Some scope delimitations should also be made explicit. The exclusion of moderators (gender, age, experience, voluntariness) as a dedicated framework dimension was a deliberate decision: moderators represent internal variables of the models, belonging to the post-selection operationalization stage. This decision does not eliminate the ability to differentiate the models, although part of this discrimination occurs indirectly, particularly through D5. Operational considerations such as sample size, analytical method, and cross-cultural validation are also not part of dimensions D1 to D5, as they belong to later stages of the research design. However, these conditions may influence the practical feasibility of the choice: models with a greater number of constructs and moderators tend to require larger samples and more complex analytical techniques, which the researcher should weigh after the selection guided by the framework. The two demonstration scenarios share the same adoption stage in D3 (initial acceptance), which did not allow demonstrating the discriminatory capacity of this dimension in the application carried out.

\section{Conclusion}\label{sec:Conclusao}

This study addressed the proposed research question by constructing a decision framework that organizes the selection of technology acceptance and use models (\ac{TAM}, \ac{TAM2}, \ac{TAM3}, \ac{UTAUT}, and \ac{UTAUT2}) through five analytical dimensions, 16 guiding questions, and descriptive suitability profiles. The artifact offers researchers a structured and conceptually grounded path for justifying model choice, shifting the decision from a frequently implicit practice to an explicit and traceable process. The demonstration in two contrasting scenarios illustrated its ability both to identify misalignments between the declared model and the actual operationalization and to support theoretically grounded choices consistent with the study conditions. Beyond the framework itself, the study makes available, as supplementary material, the structured comparative analysis of the models, the individual extraction sheets, and the expanded set of guiding questions, which can be used independently as a reference for researchers in the \ac{HCI} community and related fields.

As directions for future work, three priorities are identified. The first consists of empirically validating the framework with researchers engaged in an actual theoretical selection process, through expert studies or Delphi panels, which would allow assessing the practical usefulness and clarity of the guiding questions. The second concerns extending the scope to models from other theoretical traditions, such as Task-Technology Fit, continuance models, and emerging approaches aimed at the acceptance of systems based on \ac{AI} and \ac{LLM}, thereby broadening the coverage of the artifact. The third involves the development of tools to support the use of the framework, such as interactive guides or digital instruments that reduce dependence on interpretive judgment and facilitate adoption by researchers with varying levels of familiarity with the models in this lineage.

\section*{Supplementary Material}

The supplementary material for this study is available in a public repository \cite{anonymous_2026_19714800} and comprises three artifacts: the expanded set of guiding questions, with the 16 questions distributed across the five analytical dimensions and their respective response alternatives; the complete descriptive profiles of the five models, detailed by dimension; and the individual extraction sheets for each model, containing constructs, causal relationships, moderators, original context, and primary adoption stage.

\section*{Statement on the Use of Generative AI}

Generative artificial intelligence tools (ChatGPT, by OpenAI, and Claude, by Anthropic) were used to assist with language revision and the verification of conceptual consistency across sections of this manuscript.

\bibliographystyle{sbc}
\bibliography{Referencias/referencias}

\end{document}